\documentclass{nature_modified_arxiv}
\usepackage{graphicx}
\usepackage{hyperref}
\usepackage{marvosym}
\usepackage{multibib}
\usepackage{amssymb,amsmath}
\newcites{Methods}{References}

\def\omeg{\omega}
\def\vomeg{\boldsymbol{\omega}}
\def\AngMom{\boldsymbol{L}}
\def\Torque{\boldsymbol{N}}
\def\Inertia{\mathcal{I}}
\def\meanI{\Inertia_\mathrm{s}}
\def\dInertia{\mathcal{C}}
\def\dInertiaT{\boldsymbol{\hat{\mathcal{C}}}}
\def\InertiaT{\boldsymbol{\mathcal{\hat{I}}}}
\def\opLove{\hat{k}}
\def\kTf{k_T^f}
\def\kT{k_T}
\def\gyr{\gamma}
\def\rmf{\mathrm{rmf}}
\def\Rp{R}
\def\Mp{M}
\def\Tp{T_\mathrm{BB}}
\def\g{g}

\def\Rs{R_\star}
\def\Ms{M_\star}
\def\Ts{T_\star}
\def\sma{a}
\def\porb{P_\mathrm{orb}}

\def\Gm{G}
\def\sigmas{\sigma_\mathrm{sb}}
\def\Rearth{R_\oplus}

\def\layer{\delta}
\def\visc{\eta}
\def\convvisc{\eta_\mathrm{conv}}
\def\diff{\kappa}
\def\expansion{\alpha}
\def\dT{\Delta T}
\def\drho{\rho_{\star}}
\def\Young{\mathcal{E}}
\def\depth{d}
\def\Elast{\epsilon}
\def\Poisson{\nu}
\def\comp{c}
\def\rhom{\bar{\rho}}
\def\density{\rho_\mathrm{m}}

\def\tconv{\tau_\mathrm{c}}
\def\ttpw{\tau_\mathrm{TPW}}
\def\tr{\tau_\mathrm{R}}
\def\Xtpw{X_\mathrm{TPW}}

\def\d{\mathrm{d}}

\newcommand{\balign}[1]{
\begin{align}
#1
\end{align}}

\newcommand{\eq}[1]{Eq.\,(\ref{#1})}

\newcommand{\fig}[1]{Fig.\,\ref{#1}}

\newcommand{\tab}[1]{Table\,\ref{#1}}

\newcommand{\dd}[2]{\frac{\mathrm{d} \!\! \ #1}{\mathrm{d}\!\! \ #2}}

\title{Continuous reorientation of synchronous terrestrial planets controlled by mantle convection}

\author{J\'er\'emy Leconte$^{1}$
}
\def\jnl@style{\it}
\def\aaref@jnl#1{{\jnl@style#1}}

\def\aaref@jnl#1{{\jnl@style#1}}

\def\aj{\aaref@jnl{AJ}}                   
\def\araa{\aaref@jnl{ARA\&A}}             
\def\apj{\aaref@jnl{ApJ}}                 
\def\icarus{\aaref@jnl{Icarus}}                 
\def\apjl{\aaref@jnl{ApJ}}                
\def\apjs{\aaref@jnl{ApJS}}               
\def\ao{\aaref@jnl{Appl.~Opt.}}           
\def\apss{\aaref@jnl{Ap\&SS}}             
\def\aap{\aaref@jnl{A\&A}}                
\def\aapr{\aaref@jnl{A\&A~Rev.}}          
\def\aaps{\aaref@jnl{A\&AS}}              
\def\azh{\aaref@jnl{AZh}}                 
\def\baas{\aaref@jnl{BAAS}}               
\def\jrasc{\aaref@jnl{JRASC}}             
\def\memras{\aaref@jnl{MmRAS}}            
\def\mnras{\aaref@jnl{MNRAS}}             
\def\pra{\aaref@jnl{Phys.~Rev.~A}}        
\def\prb{\aaref@jnl{Phys.~Rev.~B}}        
\def\prc{\aaref@jnl{Phys.~Rev.~C}}        
\def\prd{\aaref@jnl{Phys.~Rev.~D}}        
\def\pre{\aaref@jnl{Phys.~Rev.~E}}        
\def\prl{\aaref@jnl{Phys.~Rev.~Lett.}}    
\def\pasp{\aaref@jnl{PASP}}               
\def\pasj{\aaref@jnl{PASJ}}               
\def\qjras{\aaref@jnl{QJRAS}}             
\def\skytel{\aaref@jnl{S\&T}}             
\def\solphys{\aaref@jnl{Sol.~Phys.}}      
\def\sovast{\aaref@jnl{Soviet~Ast.}}      
\def\ssr{\aaref@jnl{Space~Sci.~Rev.}}     
\def\zap{\aaref@jnl{ZAp}}                 
\def\nat{\aaref@jnl{Nature}}              
\def\iaucirc{\aaref@jnl{IAU~Circ.}}       
\def\aplett{\aaref@jnl{Astrophys.~Lett.}} 
\def\apspr{\aaref@jnl{Astrophys.~Space~Phys.~Res.}}
\def\bain{\aaref@jnl{Bull.~Astron.~Inst.~Netherlands}} 
\def\fcp{\aaref@jnl{Fund.~Cosmic~Phys.}}  
\def\gca{\aaref@jnl{Geochim.~Cosmochim.~Acta}}   
\def\grl{\aaref@jnl{Geophys.~Res.~Lett.}} 
\def\jcp{\aaref@jnl{J.~Chem.~Phys.}}      
\def\jgr{\aaref@jnl{J.~Geophys.~Res.}}    
\def\jqsrt{\aaref@jnl{J.~Quant.~Spec.~Radiat.~Transf.}}
\def\memsai{\aaref@jnl{Mem.~Soc.~Astron.~Italiana}}
\def\nphysa{\aaref@jnl{Nucl.~Phys.~A}}   
\def\physrep{\aaref@jnl{Phys.~Rep.}}   
\def\physscr{\aaref@jnl{Phys.~Scr}}   
\def\planss{\aaref@jnl{Planet.~Space~Sci.}}   
\def\procspie{\aaref@jnl{Proc.~SPIE}}   

\begin{document}

\maketitle

\begin{affiliations}

\item Laboratoire d'astrophysique de Bordeaux, Univ. Bordeaux, CNRS, B18N, allée Geoffroy Saint-Hilaire, 33615 Pessac, France.
\end{affiliations}


\begin{abstract}
A large fraction of known rocky exoplanets are expected to have been spun-down to a state of synchronous rotation, including temperate ones. Studies about the atmospheric and surface processes occurring on such planets thus assume that the day/night sides are fixed with respect to the surface over geological timescales. Here we show that this should not be the case for many synchronous exoplanets. This is due to True Polar Wander (TPW), a well known process occurring on Earth and in the Solar System that can reorient a planet without changing the orientation of its rotational angular momentum with respect to an inertial reference frame. As on Earth, convection in the mantle of rocky exoplanets should continuously distort their inertia tensor, causing reorientation. Moreover, we show that this reorientation is made very efficient by the slower rotation rate of synchronous planets. This is due to the weakness of their combined rotational/tidal bulge---the main stabilizing factor limiting TPW. Stabilization by an elastic lithosphere is also shown to be inefficient. We thus expect the axes of smallest and largest moment of inertia to change continuously over time but to remain closely aligned with the star-planet and orbital axes, respectively. 
\end{abstract}


\vspace{12pt}
Long before their discovery, it was hypothesized that many exoplanets would be close enough from their star to undergo tidal synchronization\cite{Dol64}. This was supported by the synchronous rotation of all major Solar System satellites, including the Moon.

This has quite dramatic implications for the planetary climate. Because one hemisphere never receives any light from the star, it has been argued that this \textit{night-side} could completely trap volatiles such as water\cite{LFC13,Men13,TLS16}, carbon dioxide\cite{Wor15,TFL17}, or even the whole atmosphere\cite{HMT96,HK12}. But the amount of volatiles that can be trapped depends crucially on various parameters -- land/ocean distribution on the day-side, topography\cite{TLS16}, geothermal heat flux below ice caps\cite{TFL17}, etc. -- that themselves depend on the planetary orientation.
Maybe more importantly, the efficiency of the Carbonate-Silicate cycle\cite{WHK81}, which may control the potential presence of liquid water on numerous temperate-to-cold planets around low-mass stars, strongly depends on the insolation and precipitations over continents. Whether the substellar point lies above a continent or a large ocean\cite{EKP12}, and whether it changes over the course of the planet's lifetime implies major changes for the atmospheric content and it stability\cite{KGM11}.

To understand the atmospheric and surface processes at play on tidally spun-down planets, it is crucial to know \textit{not only} whether they are in what we usually call a synchronous spin state or not, but also if the orientation of their surface is truly fixed with respect to their star over geological timescales!\footnote{We do not discuss here the small librations of the position of the substellar point that would be caused by the eccentricity of the orbit or a small obliquity.}

Concerning first question, it has been shown that the processes that keep Mercury and Venus out of a synchronicity could be at play on planets with eccentric orbits or having a dense enough atmosphere\cite{MBE12,LWM15}. But this still leaves us expecting plenty of synchronized close-in exoplanets.

Our goal is to address the second question that can be recast as follows: does a planet that has been tidally synchronized always shows the same face to its star? Indeed, all the studies on the rotation of exoplanets have focused on the evolution of the planetary angular velocity vector ($\vomeg$). However, as is well known in solid mechanics, the axis of rotation of a solid body usually changes with respect to this body. In other words, in a frame rotating with this fixed axis of rotation and with an angular velocity $\omeg\equiv ||\vomeg ||$ (in our case the frame where the star is fixed), the orientation of the solid (in our case the planet) can change over time. 

This process, called True Polar Wander (TPW), has happened and is still happening on Earth, as evidenced by both geological and historical records\cite{MM60,Eva03}. These records show that Earth's rotation axis has undergone large excursions (possibly up to 90$^\circ$) with respect to the planet over geological timescales, thus changing continuously which parts of the surface were receiving more (equator) or less (poles) insolation. Although, on Earth, TPW is generally accompanied by the motion of plates with respect to the mantle, plate tectonics is not needed for TPW to occur. Indeed, reorientation is also observed on many Solar System bodies\cite{MNM14}. This is crucial as many exoplanets may not be subjected to plate tectonics. Finally, note that TPW has nothing to do with the precession-nutation of the rotation axis with respect to an inertial reference frame\cite{NL97} that entails exchange of angular momentum and only add its effects to the aforementioned one. 

Here, to assess the presence of TPW on synchronous planets, we will apply a simple formalism developed for the Earth which will allow us to identify the key processes and parameters involved. Then we will show how the various stabilization mechanisms ought to be inefficient on known rocky exoplanets. It results that even a weak convection in their mantle could create a deformation sufficient to continuously reorient the planet as the convection pattern evolves. 

\section{Dynamics of True Polar Wander}


Because of heterogeneities in their interior and deformation due to rotation and tides, planets are not truly spherically symmetric bodies. The simplest, general form that the inertia tensor, $\InertiaT$, can take is thus
\balign{\InertiaT=\left(\begin{array}{ccc}A & 0 & 0 \\0 & B & 0 \\0 & 0 & C\end{array}\right)}
where the three principal moments of inertia verify $A<B<C$. The axes diagonalizing the inertia tensor are the axes of figure which define a frame attached to the solid planet and rotating with it. In this frame, the conservation of angular angular momentum, $\AngMom\equiv\InertiaT\cdot \vomeg$, yields the Liouville equation
\balign{\dd{\ }{t}\InertiaT\cdot \vomeg +\vomeg \times\left(\InertiaT\cdot \vomeg\right) = \Torque,
}
where $\Torque$ is the external torque. This equation shows that even without any torque, if $\vomeg$ is not collinear to one of the figure axes, the rotation axis will have to \textit{wander} with time\cite{GPS02}. 

Solving this equation is not trivial. However the end state of the torque free motion of a viscous planet can be determined by simply minimizing the energy
\balign{E=\frac{1}{2}\,\vomeg \cdot\InertiaT\cdot \vomeg \equiv \frac{ ||\AngMom||^2 }{2 \Inertia},
}
at constant angular momentum. The lower energy state thus corresponds to a rotation about the largest moment of inertia, i.e. where the moment of inertia about the instantaneous rotation axis $\Inertia\equiv \vomeg \cdot\InertiaT\cdot \vomeg /||\vomeg||^2$ equals $C$, as observed for all Solar System planets. For a synchronously rotating body, the tidal potential further aligns the axis of smallest moment of inertia, $A$, with the tide raising body, as is observed for all major Solar System moons.


However, in determining these axes of largest and smallest moments of inertia, one should remove the instantaneous rotational and tidal deformations\cite{Gol55,MM60,MN07}. This means that the elements of the inertia tensor should first be decomposed as follows\cite{MN07}
\balign{\Inertia_{i,j}&=\meanI \delta_{i,j}+\dInertia_{i,j}\nonumber\\
&+\opLove \frac{\Rp^5\omeg^2}{\Gm}\left[\frac{1}{3}\left(m_i m_j-\frac{1}{3} \delta_{i,j}\right)-\left(e_i e_j-\frac{1}{3} \delta_{i,j}\right)\right],
\label{inertiatensor}}
where $\meanI$ is the spherically symmetric component of the moment of inertia, $\opLove$ the Love operator\cite{MM60}, $\boldsymbol{e}$ the unit vector directed toward the tide raising body, $\boldsymbol{m}\equiv\vomeg/\omeg$, $\Rp$ the mean planetary radius, $\delta_{i,j}$ the Kronecker symbol, and $\Gm$ the gravitational constant. The last term accounts for induced deformation and only affects characteristic TPW timescale ($\ttpw$)\cite{TS07}. 
Thus, the axes of largest and smallest moments of inertia should be determined by diagonalizing the intrinsic deformation tensor caused by all non-hydrostatic effects ($\dInertiaT$).

\section{True Polar Wander and convective cycles on Earth}

Many processes cause intrinsic deformations but evidence from the gravity field show that, on Earth, the deformation is dominated by mantle convection\cite{HCR85,TS07}. Hot rising plumes create negative density anomalies but cause surface uplift, which results in net positive gravity anomalies\cite{HCR85,ZZL07}. Cold downwellings have the opposite effect.
Interestingly, this convection pattern evolves over time\cite{ZZL07} with a characteristic convective timescale ($\tconv$), and so does $\dInertiaT$.

\begin{figure}
\centering
\resizebox{0.45\hsize}{!}{\includegraphics{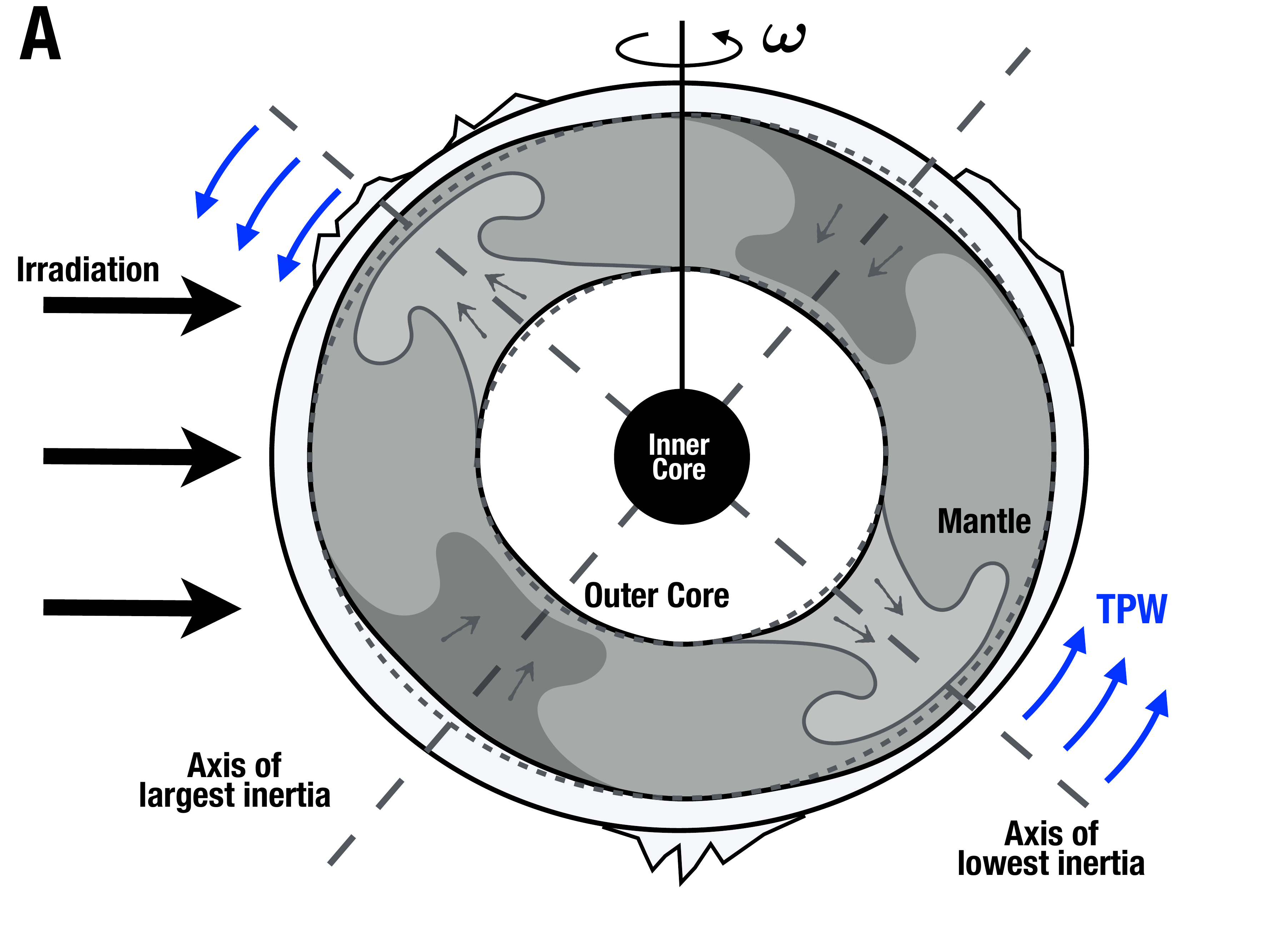}}
 \resizebox{0.45\hsize}{!}{\includegraphics{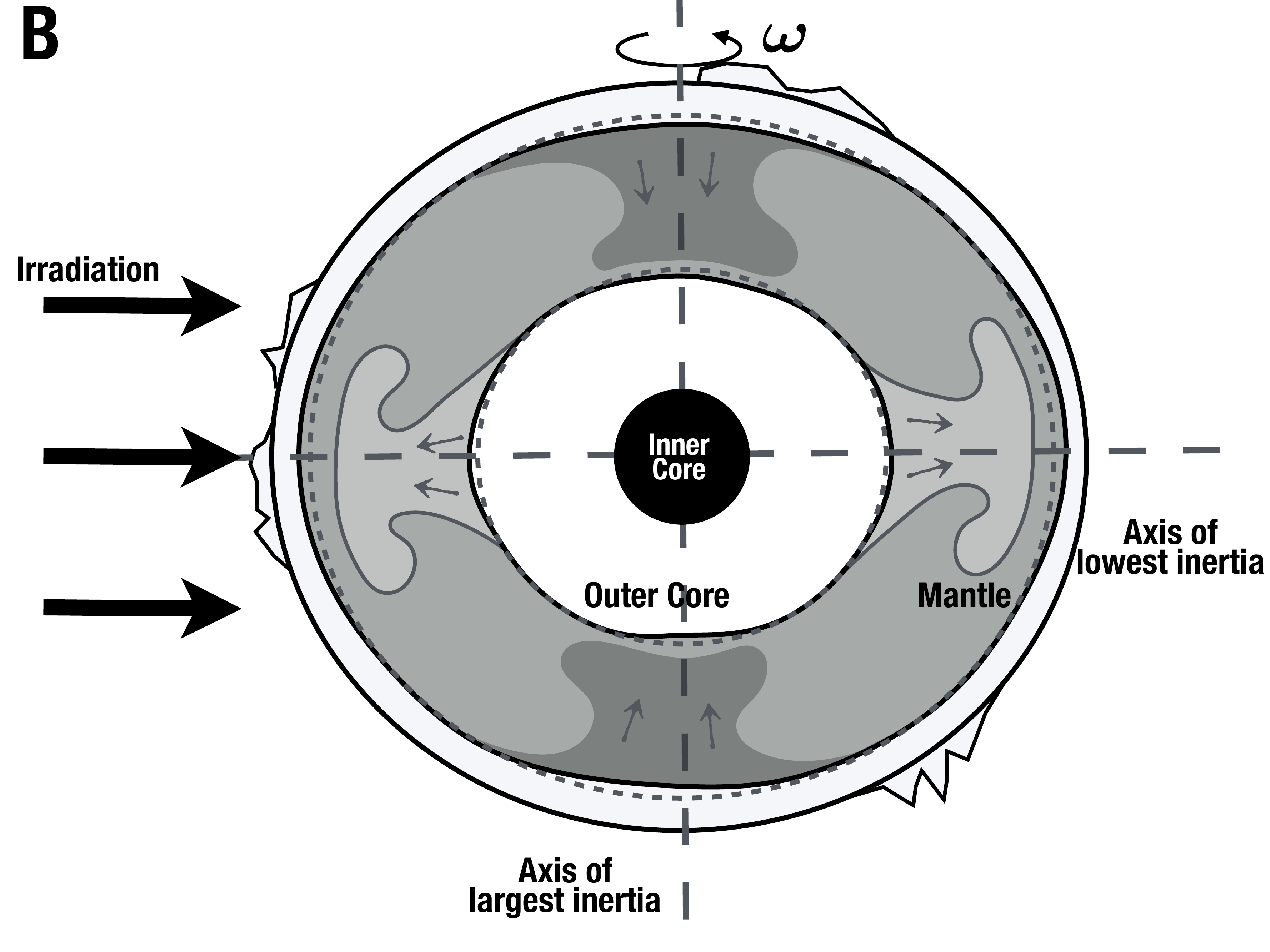}}
\caption{Schematic picture of True Polar Wander driven by mantle convection on a synchronous planet. \textbf{A}: Hot, low density (lighter shading) upwelling plumes rise and cause surface uplift. The net effect is a positive geoid/mass anomaly\cite{HCR85} (exaggerated here) that coincides with the axis of the smallest moment of inertia. In contrast, cold downwellings (darker shading) are negative anomalies where the axis of largest moment of inertia will lie. \textbf{B}: TPW will tend to align these axes with the star-planet and rotation axes, respectively. If numerous plumes are present instead of the 2-cell pattern shown\cite{Mon91}, the principal axes will be determined by the resulting degree 2 moment. Surface topography follows the reorientation. If plate tectonics occurs, continents will undergo an additional drift with respect to the mantle. }
\label{fig:schema}
\end{figure}

So the rotation axis of a planet should follow the axis of largest inertia created by mantle convection, as observed in Earth geological records\cite{Eva03}. In particular, major recorded TPW events seem to be due to the formation of hot upwelling plumes\cite{HCR85,ZZL07} which are driven to the equator (See \fig{fig:schema}). Although this process seems to be linked to the cycle of aggregation and dispersal of supercontinents\cite{Eva03,Gur88}, and may be affected by subduction\cite{GB14}, plate tectonics in general is not necessary for mantle convection to undergo cycles which affect the inertia tensor\cite{ZZL07}. 

This qualitative behavior follows well the quantitative theory outlined above. Following Tsai \& Stevenson\cite{TS07}, one can construct a figure of merit, $\Xtpw$, estimating the ability of a planet to undergo TPW. Then, the maximum TPW reached for a forcing of period $\tau$ (hereafter equal to convective period $\tconv$) is
\balign{\mathrm{TPW}_\mathrm{max}=\tan ^{-1}\left[ \sinh \Xtpw\right],\label{maxTPW}}
where
\balign{\Xtpw\equiv\frac{1}{2\pi}\frac{\langle\dInertiaT\rangle}{C-A}\frac{\tconv}{\tr}=\left(\frac{3 \gyr}{2\pi \kTf}\right)\frac{\langle\dInertiaT\rangle}{\meanI}\frac{\Gm \Mp}{\Rp^3 \omeg^2}\frac{\tconv}{\tr}. \label{TPWefficiency}}
$\tr=(19\visc)/(2\density \g \Rp)$, $\Mp$, $\g$, $\kTf$, and $\gyr=\meanI/(\Mp \Rp^2)$ are respectively the effective viscous relaxation timescale, mass, gravity, fluid Love number, and inertia factor of the planet, and $\visc$ and $\density$ the effective viscosity and density of its mantle. 

$\Xtpw$ is the ratio of the characteristic amplitude of the driving non-hydrostatic inertia anomaly ($\langle\dInertiaT\rangle$) times the forcing timescale over the stabilizing hydrostatic rotational bulge ($C-A=\meanI \times(\kTf/3\gyr)\times(\Rp^3 \omeg^2/\Gm \Mp)$) times the relaxation timescale. When $\Xtpw\ll1$, TPW$_\mathrm{max}$ is small ($\sim\Xtpw$). When $\Xtpw\gtrsim3$, the pole can shift by 90$^\circ$ during a single convective cycle. 
On Earth, the convective timescale and amplitude seem to be on the order\cite{TS07} of $\tconv\sim 100$\,Myr, and $\langle\dInertiaT\rangle/\meanI\sim 10^{-5}$. Using the parameters from \tab{tab:params} yields $\tr\approx3\times 10^{4}$\,yr and $\Xtpw\approx 1-2$, meaning that large TPW events can arise over timescales $\gtrsim$100\,Myr, but rotation drastically filters out shorter events.

\section{The case of synchronous terrestrial exoplanets}

What is the efficiency of TPW on synchronous planets? Of course, very little is known about these objects beyond a few global parameters. Fortunately our analysis reveals that the most important parameter is the rotation period, which is equal to the orbital one for a synchronous planet.

Indeed, simple boundary layer scaling arguments predict that $\tconv\sim \layer^2/\diff$, where $\layer=5(\convvisc \diff/\density \g \expansion \dT)^{1/3}$ is the boundary layer depth, $\diff$ the thermal diffusivity, $\convvisc$ the effective viscosity for convection, $\expansion$ the thermal expansion coefficient and $\dT$ the temperature difference driving the convection. The convective inertia anomaly can be roughly estimated using $\langle\dInertiaT\rangle_\mathrm{conv}/\meanI \sim0.1\expansion \dT \layer/\Rp$ which gives reasonable values for Earth\cite{TS07}. 

Interestingly, combining these scaling yields
\balign{\Xtpw\approx0.2\,\frac{\convvisc}{\visc}\frac{\Gm \Mp}{\Rp^3 \omeg^2},}
where all internal parameters have disappeared except for the ratio $\convvisc/\visc$ ($\approx$1/30 for the Earth\cite{TS07}; see Methods).
So TPW efficiency is determined by the the viscosity structure and not the absolute viscosity\cite{HCR85, TS07}. Global variations in viscosity due to a hotter/colder interiors should not substantially affect our results, which only depend directly on measurable quantities.

\begin{figure}
\centering
\resizebox{0.65\hsize}{!}{\includegraphics{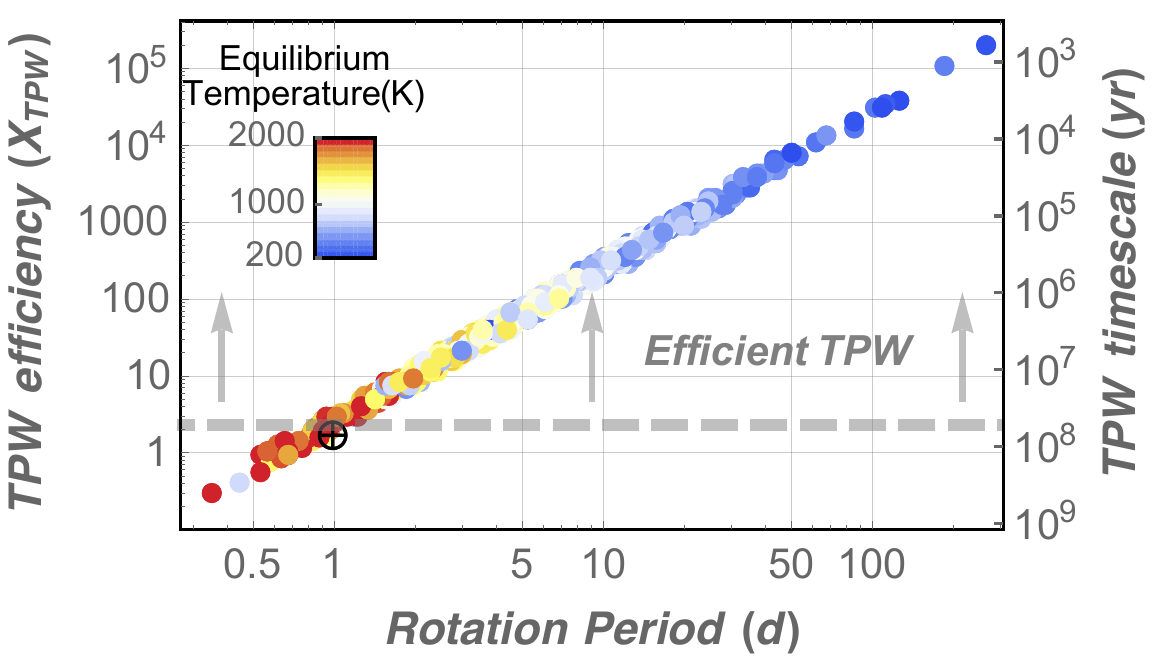}}
\vspace{-0.cm}
 \caption{Efficiency of true polar wander ($\Xtpw$) on known rocky exoplanets as a function of their orbital period (dots). The $\oplus$ symbol shows the value of $\Xtpw$ for the Earth. The color of the dots refers to the equilibrium blackbody temperature of the planet ($\Tp$) determined assuming a complete redistribution of the incoming stellar energy (See methods). As expected, cooler planets have longer orbital periods. All synchronous planets with an orbital period above 1-2\,days should undergo true polar wander very easily. The right ordinate axis shows the timescale at which the pole is able to follow the axis of largest moment of inertia ($\ttpw$; See methods).}
 \label{fig:Xtpw}
\end{figure}

\begin{figure}
\centering
\resizebox{0.6\hsize}{!}{\includegraphics{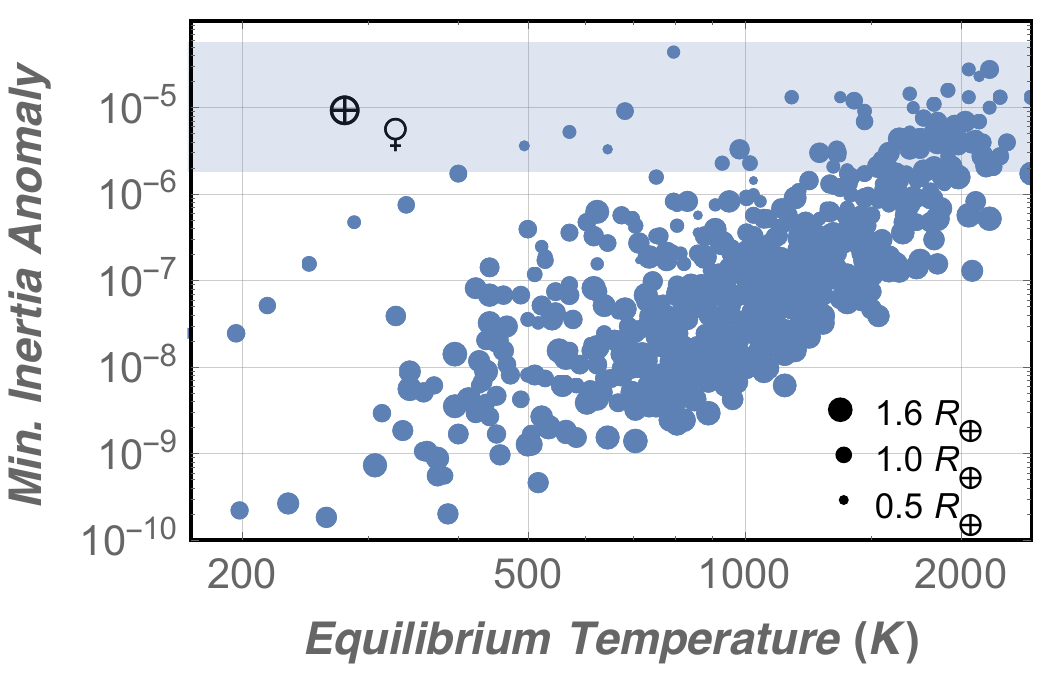}}
 \vspace{-0.cm}
\caption{Minimal inertia anomaly ($\langle\dInertiaT\rangle_\mathrm{min}/\meanI$) needed to excite significant polar wander as a function of the planetary temperature ($\Tp$). The size of the dot is proportional to the size of the planet. The $\oplus$ and \Female \ symbols show the convective inertia anomaly for the Earth and Venus. The shaded area illustrates the range of convective contribution predicted by our simple scaling by varying the radius of the planet between 0.5 and 1.6\,$\Rearth$ and the convective viscosity within two orders of magnitude. For temperate planets, mantle convection would have to be two to three orders of magnitude less vigorous than on Earth to suppress TPW.}
 \label{fig:ctpw}
\end{figure}

\fig{fig:Xtpw} shows the TPW efficiency for all known transiting planets with $R<1.6\,\Rearth$ to select terrestrial bodies\cite{Rog15}. As anticipated, rotation rate is the determining factor, planetary density not varying much in the range of radii considered.
It demonstrates that planets with an orbital period $\gtrsim$1-2\,days offer very favorable conditions for TPW, like Solar System moons\cite{MNM14}. Because the TPW timescale is shorter than the convective one, we can expect the reorientation to follow the convective pattern quasi-statically.

One limitation of our calculation is that our convective contribution to the inertia anomaly is scaled on the Earth, where plate tectonics is strongly coupled to mantle convection\cite{Gur88}. While there would be no continental motions on a planet without plate tectonics, there is no reason that convection could not distort its crust, or simply cause sufficient density heterogeneities within the mantle. To quantify this we computed the minimal inertia anomaly that would have to be created by mantle convection to excite TPW (See methods). \fig{fig:ctpw} shows that the absence of moving plates would have to reduce the distortion by two to three orders of magnitude to suppress TPW which seems unlikely. Indeed Venus' triaxiality is only about a factor two smaller than the Earth one despite its absence of plate tectonics. 

Another impediment to a truly continuous reorientation would be a lithosphere with a permanent elasticity\cite{MN07,Wil84}. 
For Mars, the stabilizing effect of the remnant rotational bulge inherited from the solidification of the lithosphere plays an important role in the TPW event that followed the formation of Tharsis\cite{DMM08,Wil84}.
The lithosphere being strong enough to support part of the Tharsis load once at the equator, the load itself became a stabilizing factor against subsequent TPW events, strongly decoupling the motion of the pole from the convective cycle.


Do we expect an equivalent stabilization by the lithosphere of exoplanets? Importantly, Earth's lithosphere does not seem able to support such loads permanently\cite{TWH81}. This is demonstrated by the absence of a geoid signature of a remnant rotational bulge inherited from a faster, past rotation and by the weakness of the topography to geoid correlation at long wavelengths. This difference is easily explained by estimating the dimensionless rigidity of a spherical lithosphere due to membrane stresses\cite{TWH81,WT82}
\balign{\Elast=\frac{\Young \depth}{\g \Rp^2\drho},\label{elasticity}}
where $\Young$ and $\depth$ are the Young modulus and thickness of the lithosphere, and $\drho$ a density characterizing the load. 
Decreasing with planetary mass, the rigidity is sufficient to support massives loads on Mars ($\Elast\sim0.5$), but not on Earth ($\Elast\sim0.02$)\cite{TWH81}, and even less on larger planets

\begin{figure}
\centering
\resizebox{0.5\hsize}{!}{\includegraphics{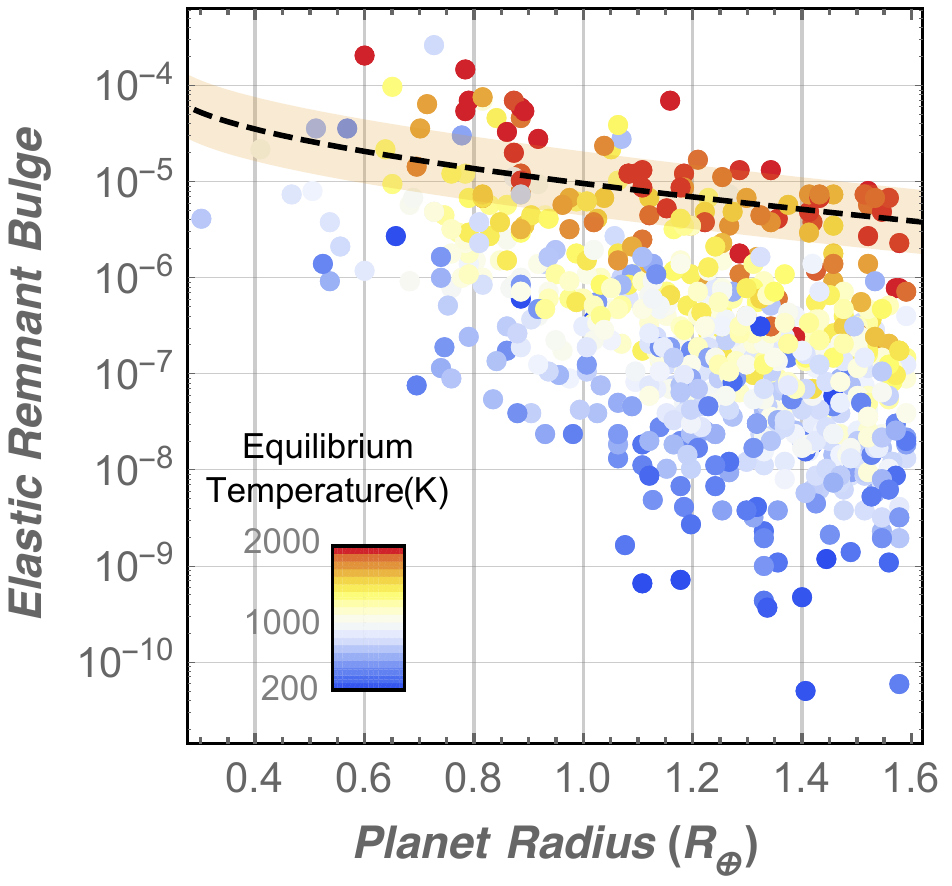}}
 \caption{Dimensionless contribution of the elastic remnant bulge to the inertia deformation tensor as a function of the planet radius for all known rocky exoplanets ($\langle\dInertiaT\rangle_\mathrm{lit}/\meanI $; See methods). The color of the dot shows the equilibrium temperature of the planet. For comparison, the dashed black line shows the contribution of convective motions to the inertia tensor expected from our simple scaling ($\langle\dInertiaT\rangle_\mathrm{conv}/\meanI $; See text). The shaded area illustrates the uncertainty on this prediction by varying the convective viscosity within two orders of magnitude. For all warm and temperate planets, the deformation is expected to be dominated by convective motions.}
\label{fig:elast}
\end{figure}

The effect of an elastic lithosphere is assessed in \fig{fig:elast}. It shows the contribution of a potential remnant rotational/tidal bulge frozen-in during the formation of the lithosphere\cite{WT82,Wil84,MN07,MNM14} (See methods), and how it compares to the convective contribution. Except for the hottest planets, the elastic bulge is much weaker than the convective one. This results from both the low rotation rate of these objects and their lower rigidity. For the same reason, the stabilizing effect of a topographic load such as Tharsis would be reduced by orders of magnitude.

Our findings collectively suggest that reorientation of warm to temperate rocky exoplanets should be continuous and controlled by mantle convection, as on Earth. Hence, it should still occur today, except for the smallest bodies for which interior cooling has been fast enough to shut down convection, although tidal heating can be a substantial source of geothermal heat for close-in planets. 






\section{Conclusion}


There is ample evidence that Earth underwent several episodes of large TPW. It is thus sensible to assume that many rocky exoplanets undergo TPW as well. Our analysis reveals that synchronous planets should exhibit efficient TPW because the stabilization by their hydrostatic and elastic bulges is weak. Interestingly, if plate tectonics does occur on these planets, we predict that supercontinents should be either at the sub-stellar or anti-stellar point -- depending on the initial conditions -- during their formation. Facilitating the formation of hot upwellings beneath them\cite{Gur88,ZZL07}, supercontinent are indeed expected to sit on the axis of smallest moment of inertia\cite{HCR85,Eva03} which should align with the star. 

Knowing that, it would be interesting to quantify how the partitioning of volatiles between the atmosphere, surface, and mantle varies over time. By simply affecting the day side ocean/land fraction, TPW periodically changes, among other things, the heat redistribution efficiency of the atmosphere and ocean toward the night side, the weathering efficiency\cite{EKP12,KGM11}, and thus the planetary global greenhouse gas content, temperature, and climate. Changes in the orientation of the large scale topography and geologically active regions may play a role as well\cite{TLS16,TFL17}. Generally, when studying geological processes linked to the orientation of the planet, the carbonate-silicate cycle being one example\cite{EKP12}, one should consider that this orientation can evolve as fast or faster than the processes studied.


\vspace*{30pt}
\bibliographystyle{naturemag}
\bibliography{biblio.bib}


\begin{addendum}

\item[Acknowledgements] 

This project has received funding from the European Research Council (ERC) under the European Union's Horizon 2020 research and innovation program (grant agreement No. 679030/WHIPLASH). The author thanks E. Couderc, R. Jolivet, L. Londeix, and F. Selsis for comments on the initial manuscript. 

\item[Author contributions] 

\item[Correspondence] Correspondence and requests for materials should be addressed to J.L.

\item[Competing interests] The authors declare that they have no competing financial interests.

\end{addendum}

\begin{table}
\caption{Nominal parameters used in numerical estimates.\label{tab:params}}
\centering
\begin{tabular}{llll}
Name & Symbol & Value & unit\\
\hline
Viscosity & $ \visc $  &  $3\cdot 10^{22}$&  Pa.s\\
Convective viscosity & $ \convvisc  $  & $1\cdot 10^{21}$& Pa.s\\
Mantle density & $ \density  $  & $4000$& kg/m$^{3}$\\
Thermal diffusivity & $\diff$  &$1\cdot 10^{-6}$ & m$^{2}$/s \\
Thermal expansion & $\expansion  $  &$2\cdot 10^{-5}$ & K$^{-1}$ \\
Temperature difference & $\dT$ & 500 & K\\
Fluid Love number & $ \kTf $  & 1& \\
Effective tidal Love number & $ \kT $  & 0.3 & \\
Reduced moment of inertia & $ \gyr $  & 1/3 & \\
Lithospheric thickness & $ \depth $  & 50 & km\\
Young Modulus & $ \Young $  & $6.5\cdot 10^{10}$ & Pa\\
Poisson ratio & $ \Poisson $  & 0.25 & \\
 \hline\\
\multicolumn{4}{l}{Parameters taken from refs.\cite{TS07,TWH81}}\\
\end{tabular}
\end{table}

\pagebreak


\clearpage
\begin{methods}

\subsection{Mass-Radius relationship.}

The mass of rocky bodies discovered by transit surveys is often difficult to measure. To infer it from a given measured radius, we use the mass radius relationship from ref.\cite{FMB07} which reads
\balign{
\Rp&=\left(0.3102\,\rmf +0.7932\right) +\left(0.2337 \,\rmf +0.4938\right) \log_{10} \Mp\nonumber\\
&+\left(0.0592 \,\rmf +0.0975\right) \left(\log_{10} \Mp\right)^2,
}
where $\rmf$ is the rock mass fraction taken to be 0.67 as for the Earth. Being a degree 2 polynomial in $\log_{10} \Mp$, this equation can be inverted analytically. 

\subsection{Definition of the black body temperature scale.}

 For most known transiting planets, the exoplanets.org catalog provides the stellar radius, mass, and effective temperature, along with the orbital semi-major axis (respectively $\Rs$, $\Ms$, $\Ts$, and $\sma$). The total power received by the planet is thus given by
\balign{L_\mathrm{p}=\sigmas \Ts^{4} \left(\frac{\Rs}{\sma}\right)^2 \pi \Rp^2,}
where $\sigmas$ is the Stefan-Boltzmann constant. As we only want a \textit{"flux temperature"}, we equate this incoming power with the power emitted by the planet if it were a black body with a uniform temperature ($\Tp$)
\balign{L_\mathrm{p}= 4\pi \Rp^2\,\sigmas \Tp^{4},}
which yields
\balign{\Tp= \frac{\Ts}{4^{1/4}}  \sqrt{\frac{\Rs}{\sma}}.}
This gives us a temperature scale more than an accurate idea about the real temperature at the surface that can be affected by both albedo and greenhouse effect. With these conventions, the equilibrium temperatures are 279\,K and 330\,K for the Earth and Venus, respectively. The fact that $\Tp$ depends strongly on the stellar type explains the huge scatter observed in the orbital period / temperatures relation (See \fig{fig:Xtpw}).

\subsection{Limit period for synchronization.}

One can have a rough estimate of the maximal distance at which planets are synchronized by tides over a time $\tau_\mathrm{tid}$ by equating the angular momentum to be removed ($\sim\meanI \omeg_0$) to the integrated tidal torque over that time. Assuming a simple constant phase lag model, one finds
\balign{\int \mathcal{N}_\mathrm{tid} \d t \approx \mathcal{N}_\mathrm{tid} \tau_\mathrm{tid}=\frac{3}{2} \frac{k_T}{Q}\frac{\Gm \Ms^2 \Rp^5}{\sma^6} \tau_\mathrm{tid},}
where $k_T$ and $Q$ is the effective tidal Love number and quality factor.
Interestingly, using Kepler third law, one can get rid of stellar parameters which yields
\balign{\meanI \omeg_0=\frac{3}{2} \frac{k_T \Rp^5}{Q}\left(\frac{ 2\pi}{\porb}\right)^2 \tau_\mathrm{tid},}
where $\porb$ is the orbital period.
This stems from the fact that the tidal and rotational potential have very similar forms when the rotation rate is replaced by the orbital mean motion. This entails that we can find the orbital period at which a planet will be synchronized independently of the stellar type. This limit orbital period below which planets are expected to be synchronous is given by
\balign{\porb^{\mathrm{sync}}(\tau_\mathrm{tid})&=2\pi\left(\frac{3}{2}\frac{\kT}{\gyr Q}\frac{\Rp^3}{\Gm \Mp}\frac{\tau_\mathrm{tid}}{\omeg_0}\right)^{\frac{1}{4}}\nonumber\\
&=245 \mathrm{d} \left(\frac{13}{Q}\frac{\tau_\mathrm{tid}}{4.5\,\mathrm{Gyr}}\right)^{\frac{1}{4}},
}
where the numerical estimate is for a planet the size of the Earth starting with a 1\,day rotation period. The $Q\approx13$ seems appropriate for Earth\cite{GS66}. As the oceans are expected to generate about 90\% of the dissipation, we can expect $Q\approx 100$ for a dry planet like Venus, lowering the critical period to $\sim$150\,days. Because of the crude dissipation model, and the assumptions needed on the initial spin, this should be regarded as a guide rather than a hard limit. Indeed, Venus as been efficiently spun down despite its $\sim$225\,day orbital period. This suggests that all the rocky planets shown in \fig{fig:Xtpw} could potentially be synchronous if no other process is at play.

\subsection{Effective viscosity and viscosity structure in the mantle}

In our analysis, we differentiate the viscosity that is supposed to be representative of convective processes ($\convvisc$) from the effective viscosity entering the relaxation timescale of the mantle ($\visc$). This might be needed as viscosity varies over several orders of magnitude throughout the mantle\cite{MF04} and different processes, being affected differently by the viscosity structure of the mantle, may thus exhibit different effective viscosities.

Empirically, a low convective viscosity on the order of $\visc\sim 10^{21}$\,Pa\,s is needed to recover a convective timescale on the order of 100\,Myr for the Earth, and is consistent with the viscosities inferred in the upper mantle\cite{MF04}. However, using this viscosity to compute the viscous relaxation timescale of the mantle yields a timescale $\sim\,100$\,yr which is very short compared to the observed timescales for the postglacial rebounds. The difference can be understood by saying that relaxation has occurred only when the slower, more viscous, deep mantle has relaxed. Following Tsai \& Stevenson\cite{TS07}, we use a value of $\visc\sim30\, \convvisc$, but note that reducing the difference between the two values here would only increase the efficiency of polar wander. 

\subsection{Efficiency of true polar wander of a Maxwell Earth and limitations.}

The ability of a planet to undergo fast true polar wander can be quantified through different means. In \fig{fig:Xtpw}, we chose to represent the dimensionless efficiency $\Xtpw$ given by Equation \ref{TPWefficiency} which quantifies the maximum polar motion over one excitation period ($\tconv$) through Equation \ref{maxTPW}. But TPW can also be seen as a low-pass filter, all perturbation with a timescale below $\ttpw$ being damped. This timescale can be defined through
\balign{\Xtpw \equiv \frac{\tconv}{\ttpw},}
yielding
\balign{\ttpw \equiv 2\pi \,\frac{C-A}{\langle\dInertiaT\rangle}\, \tr.}
It is a useful timescale as it also tells us at which rate the pole can adapt to a change in the inertia tensor. Particularly relevant here is the fact that the pole will follow the axis of maximum inertia closely if $\tconv \gg \ttpw$ (See \fig{fig:Xtpw}). 

Another metric, used in \fig{fig:ctpw}, is the minimum characteristic amplitude of the inertia deformation needed to cause a significant shift over one convective period. It can be defined as
\balign{\Xtpw \equiv \frac{\langle\dInertiaT\rangle}{\langle\dInertiaT\rangle_\mathrm{min}},}
which yields
\balign{\frac{\langle\dInertiaT\rangle_\mathrm{min}}{\meanI} \equiv 2\pi \,\frac{\tr}{\tconv}\, \frac{C-A}{\meanI}.}

In their analysis, Tsai and Stevenson\cite{TS07} disregarded a term of order $C/(\omeg \tr (C-A)),$ because it is indeed very small for the Earth. For very slow rotation rates, this term starts to be significant. Indeed it has been shown that for Venus, this term is responsible for the observed wobble of the rotation axis of about 0.5$^\circ$ around the axis of maximum moment of inertia\cite{SSB96}. For the planets with the largest orbital periods in our sample ($\gtrsim100-200$\,day), we can thus expect such small deviations from a true alignement. 

Another limitation of this analysis is that the effect of the tidal bulge is neglected, as appropriate for an application to the Earth. However we are interested only in the timescale involved and not the precise trajectory followed by the planet during TPW events. To estimate the impact of the tidal bulge on this timescale, it is important to remember that for a synchronous planet, the magnitude of the tidal and rotational bulges are about equal. This stems from the fact that the tidal bulge scales as $ \Ms \Rp^5 /\sma^3,$ which is equal to $\omeg^2 \Rp^5/\Gm$ for a synchronous planet. From Equation \ref{inertiatensor} we can therefore show that the stabilizing bulge is equal to
\balign{C-\frac{A+B}{2}= \left\{\begin{array}{l}1/3\ \mathrm{(rotation\ only)} \\5/6\ \mathrm{(including\ tides)}\end{array}\right\} \times\opLove \frac{\Rp^5\omeg^2}{\Gm}}
We thus expect the tidal bulge to impact the order of magnitude estimates presented here only by a factor of order unity which is acceptable considering the other sources of uncertainty.

\subsection{Remnant rotational bulge supported by an elastic lithosphere.}

Before a planet has cooled down sufficiently to form an elastic lithosphere, it responds to a static deforming potential hydrostatically and the final deformation is given by the last term in \eq{inertiatensor} where the Love number is taken equal to the secular, hydrostatic tidal Love number $\kTf$ which depends on the density profile alone. For an isodensity planet $\kTf=3/2$. Once a lithosphere forms, the secular Tidal Love number changes to account for its elasticity. This also creates a remnant inertia bulge supported by the lithosphere which is given by\cite{MNM14,MN07}
\balign{\Inertia_{i,j}^R&=(\kTf-\kT) \frac{\Rp^5\omeg^2}{\Gm}\left[\frac{1}{3}\left(m^*_i m^*_j-\frac{1}{3} \delta_{i,j}\right)-\left(e^*_i e^*_j-\frac{1}{3} \delta_{i,j}\right)\right].}
Here $\boldsymbol{e^*}$ and $\boldsymbol{m^*}$ are the unit vectors directed toward the tide raising body and the rotation pole at the moment of the lithosphere formation. 

The difficulty generally lies in the determination of $\kT$ that must account for the rheology of the whole planet. As we are concerned with order of magnitude estimates, we will assume that only the thin lithosphere supports the remnant bulge\cite{WT82,Wil84}. It can then be shown from equation 1-3 and 16 of ref.\cite{Wil84} that for the rotational part
\balign{\Inertia_{i,j}^R&=\frac{1-\comp}{2}\frac{\Rp^5\omeg^2}{\Gm}\left(m^*_i m^*_j-\frac{1}{3} \delta_{i,j}\right),}
where $\comp$ is the degree of compensation that measures the resistance of the lithosphere to deformation\cite{WT82,Wil84}
\balign{\comp\equiv\left[1+\frac{4 \,\Elast}{\left(5+\Poisson\right) \left(1-\frac{3 \density}{\rhom}\right)}\right]^{-1}.}
In this formula, $\Poisson$ is the Poisson ratio of the lithosphere, and $\rhom$ the mean density of the planet. The dimensionless rigidity, $\Elast$, is given by Equation \ref{elasticity}. This formula neglects the effect of bending stresses that have been shown to be unimportant for the degree-2 perturbations considered here\cite{TWH81}. It is however valid for topographic loads as well as tidal and rotational deformations as long as the right definition of $\drho$ is taken in $\Elast$. For topographic loads, $\drho$ must be equal to the density difference between the mantle and the load above. To compute the effect of rotation and tidal deformation, one must use\cite{WT82} $\drho=\density$.

By identification, this simply yields
\balign{\kTf-\kT \equiv \frac{3}{2}\left(1-\comp\right). \label{lovetocomp}}
The validity of this formula has been tested by comparing the prediction for the secular Love number of Mars to published models\cite{DMM08,SS97} for various lithospheric thicknesses. Discrepancies are found to be below the 15\% level (see Sup. \fig{fig:comp})

As a result, the characteristic amplitude of the dimensionless inertia anomaly linked to the remnant stabilizing bulge is given by\cite{Wil84}
\balign{\frac{\langle\dInertiaT\rangle_\mathrm{lit}}{\meanI} =\frac{\kTf-\kT}{3\,\gyr}\,\frac{\omeg^2 \Rp^3}{\Gm \Mp}=\frac{1-\comp}{2\,\gyr}\,\frac{\omeg^2 \Rp^3}{\Gm \Mp},}
where factors of order unity due to the geometry of the tidal and rotational bulges have been discarded. This is the quantity shown in \fig{fig:elast}. In this context, the values calculated for $\kTf-\kT$ supersede the value shown in \tab{tab:params}.

Our estimate is conservative in several ways. First, the results in \fig{fig:elast} assume a constant lithospheric thickness for all planets equal to 50\,km, which is commonly used value for the Earth\cite{TWH81}. It can be argued that this value should be smaller for larger and/or hotter planets because of the higher temperatures in the crust and mantle due to both the larger irradiation and the larger geothermal flux. Mars, for example, is observed to have a much thicker lithosphere. In addition, higher temperatures should tend to decrease the Young modulus of the lithosphere, further weakening the latter. Hence, for the hottest planets in our sample the increase in the size of the remnant bulge visible in \fig{fig:elast} could be offset by these effects. This should be further ascertained using a coupled thermal/rheological model.

\begin{figure}
\centering
\resizebox{0.5\hsize}{!}{\includegraphics{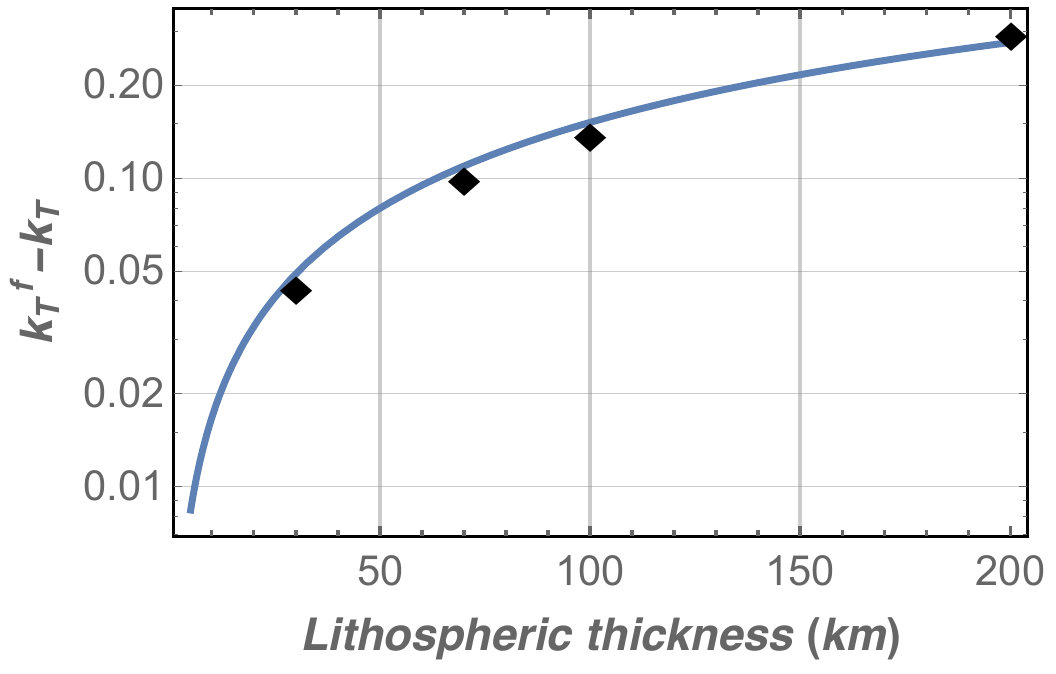}}
 \caption{Effect of lithospheric thickness on the difference between the degree-2 tidal Love numbers with ($\kT$) and without ($\kTf$) the effect of an elastic lithosphere for Mars. The black diamonds represent the calculations from Table 1 of Daradich et al.\cite{DMM08} based on a reference 5-layer Mars model\cite{SS97}. For comparison, the solid curve shows the prediction from \eq{lovetocomp} for Mars using the same numerical parameters as in the aforementioned study (Compared to the fiducial numerical values from \tab{tab:params}, only the Young Modulus was changed to $\sim$100\,GPa to be representative of the crust in the reference Mars model\cite{SS97}; see their Figure 4). Our simple analytical estimate agrees with the tabulated values to within 15\%.}
\label{fig:comp}
\end{figure}


\subsection{Code availability.}
Data processing routines are available on request from J.L. 

\subsection{Data availability.} 

Data taken from the \textit{exoplanets.org} database; retrieved on July 2017.

\end{methods}


\bibliographystyleMethods{naturemag}
\bibliographyMethods{trappist1_methods}

\end{document}